\documentclass{IEEEtran}
\usepackage{amsmath,amssymb,amsfonts}
\usepackage[pdftex]{graphicx}
\usepackage{hyperref}
\hypersetup{colorlinks, linkcolor={blue},citecolor={blue}, urlcolor={blue}}

\newcommand{\abs}[1]{\left\lvert #1 \right\lvert }
\newcommand{\modul}[1]{\left\lvert #1 \right\lvert }
\newcommand{\modulc}[1]{\left\lvert #1 \right\lvert ^{2}}
\newcommand{\aver}[1]{\left \langle #1 \right \rangle}
\newcommand{\mean}[1]{\overline{#1}}
\newcommand{\av}[1]{\langle #1 \rangle}

\newcommand{\Ntot}{N_\textrm{tot}}
\newcommand{\Npix}{N_\textrm{pix}}
\newcommand{\Nunc}{N_\textrm{unc}}
\newcommand{\rholim}{\rho_\textrm{lim}}
\newcommand{\nlim}{n_\textrm{lim}}
\newcommand{\rhobar}{\mean{\rho}_f}
\newcommand{\alim}{\alpha_\textrm{lim}}
\newcommand{\sdb}{\sigma_\textrm{dB}}
\newcommand{\Ec}{\mathcal{E}}

\usepackage{color}
 
  \usepackage{cite} % for compressed citation [1], [2] [3] -->[1]--[3]
\begin{document}
\title{Uncorrelated Configurations and Field Uniformity in Reverberation Chambers Stirred by Tunable Metasurfaces}
 
\author{
J.-B. Gros, G. Lerosey, F. Mortessagne, U. Kuhl, and O. Legrand
\thanks{Submitting date	This work was supported in part by the French ``Minist\`{e}re des Arm\'{e}es, Direction G\'{e}n\'{e}rale de l'Armement''}
\thanks{J.-B. Gros is with the Institut Langevin, CNRS UMR 7587, ESPCI Paris, PSL Research University, 75005 Paris, Franc ( e-mail: jean-baptiste.gros@espci.fr). }
\thanks{G. Lerosey is with Greenerwave, ESPCI Paris Incubator PC'up, 75005 Paris, France (e-mail: geoffroy.lerosey@greenerwave.com)}
\thanks{U. Kuhl, O. Legrand, and F. Mortessagne are with the Universit\'e C\^ote d'Azur, CNRS, INPHYNI, 06100 Nice, France,
(e-mail:ulrich.kuhl@unice.fr - olivier.legrand@unice.fr - fabrice.mortessagne@unice.fr)}
}

\maketitle

\begin{abstract}
Reverberation chambers are currently used to test electromagnetic compatibility as well as to characterize antenna efficiency, wireless devices, and MIMO systems.
The related measurements are based on statistical averages and their fluctuations.
We introduce a very efficient mode stirring process based on electronically reconfigurable metasurfaces (ERMs).
By locally changing the field boundary conditions, the ERMs allow to generate a humongous number of uncorrelated field realizations even within small reverberation chambers.
We fully experimentally characterize this stirring process by determining these uncorrelated realizations via the autocorrelation function of the transmissions.
The IEC-standard uniformity criterion parameter $\sdb$ is also investigated and reveals the performance of this stirring.
The effect of short paths on the two presented quantities is identified.
We compare the experimental results on the uniformity criterion parameter with a corresponding model based on random matrix theory and find a good agreement, where the only parameter, the modal overlap, is extracted by the quality factor.
\end{abstract}

\begin{IEEEkeywords}
Antenna characterization, reverberation chambers, metasurface, correlation function.
\end{IEEEkeywords}

\section{Introduction}

Electromagnetic (EM) compatibility extensively uses mode-stirred reverberation chambers (RC). In particular, the latter allow to obtain statistically valuable information about the electromagnetic radiation suffered by an object under test or emitted by an antenna \cite{che18,Corona2002,Holloway2003,Corona1981,Leferink2012,Amador2011,hol12,Carlsson2017,Hallbjorner2001,davy2,Besnier2014,Fiumara2005,kil04,Krauthauser2010,Krouka2018}. However, in order to get reliable statements about the fluctuations of the EM field, it is necessary to use statistically uncorrelated
experimental realizations \cite{Lunden2000,Madsen2004,Krauthauser2005,Lemoine2007,Flintoft2016,Remley2016,Gradoni2013,Oubaha2018}. In other words, the statistical ensemble usually achieved by a so-called
mode stirrer has to be mixing enough to make the intensity patterns of the chamber statistically independent from each other  \cite{Serra2017}.
As the solutions to Maxwell's equations under given boundary conditions depend continuously on the latter, a sufficiently large
change has to be performed. For mechanical mode stirrers, this usually means that the angle of rotation has to
be large enough, assuming the stirrer is not too small. Once this minimally necessary step width is determined, the number of independent samples from a full turn of the stirrer can be calculated~\cite{IEC_standard}.
In the low frequency regime, the accessible number of uncorrelated configurations may not be sufficient to meet the requirements of the International Standard IEC 61000-4-21 \cite{IEC_standard}.

In order to overcome the limitations of mechanical stirring, we propose to stir the EM field by locally changing the EM boundary conditions using electronically reconfigurable metasurfaces (ERMs) \cite{kaina2014hybridized,Kaina2015,Dupre2015,localiz,mimo,SMM_TM,Gros2019}. This can be easily implemented in any commercial reverberation chamber without any major modification. Note that the idea to use electronically reconfigurable boundary conditions to stir the EM field efficiently in a reverberation chamber was previously proposed \cite{Kingler2005,Serra2017}, but had not been experimentally exploited till now. More recently, improving the field uniformity in a reverberation chamber by means of metasurfaces \cite{Wanderlinder2017} or by making the chamber chaotic \cite{Gros2014Wamot,Gros2015aem,Gros2015,Arnaut2001,Orjubin2009,Gradoni2014,Moglie2015,Bastianelli2017,Kuhl2017} was demonstrated using mechanical or frequency stirring. In this paper, we can simultaneously stir the EM field and improve the field uniformity with the ERMs.

In this paper we focus both on the number of uncorrelated configurations determined by the field autocorrelation function and on the field uniformity characterized by the spatial fluctuations of the EM field maxima through the parameter $\sdb$. Both quantities are proposed in the standard \cite{IEC_standard} and are heavily used in the EMC community \cite{Gros2015aem,Gros2015,Lemoine2009,Moglie2015,Bastianelli2017,Madsen2004,Krauthauser2005,Krauthauser2007,Oubaha2018}. A chief assumption of the standard is the absence of unstirred components which can be related to short paths or direct processes \cite{Hart2009,Yeh2010,Gradoni2014,Corona2000,Lemoine2011,Lunden2007}. Since the latter cannot always be avoided, we address this problem as well.

In section \ref{sec:RCMeta} we describe our experimental set-up in detail including the stirring process with the ERMs.
Additionally, the effect of the short paths on the investigated complex transmissions $S_{1i}$ is shown and the centered complex transmission, a quantity free of this effect, is introduced.
In the next section, considering the latter case, we introduce the parametric autocorrelation function with respect to the stirred states related to the configurations of the ERMs.
From this we extract the number of uncorrelated realizations of the chamber.
We show that we can adjust the stirring process such that all acquired samples are uncorrelated and their number is virtually unlimited.
As a consequence, the field uniformity parameter $\sdb$ is statistically well-behaving and satisfies the requirements of the standard.
In the following section, we give an overview of the predictions concerning the field statistics based on Random Matrix Theory (RMT) and compare these predictions with our experimental results.
The effects of short paths are investigated in section~\ref{sec:short}.
We find a reduction of the number of uncorrelated realizations using the same stirring process and an increase of $\sdb$ accordingly.
We finally conclude in the last section.

\section{ Reverberation chambers with reconfigurable metasurfaces}
\label{sec:RCMeta}

\begin{figure}
	\centerline{\includegraphics[width=0.85\columnwidth]{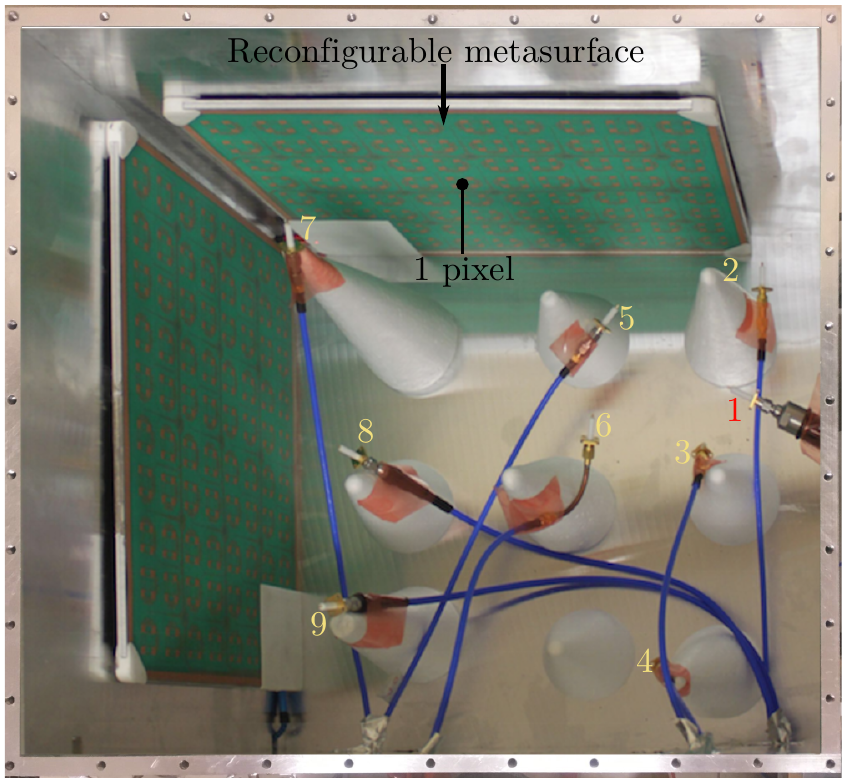}}
	\caption{\label{fig_manip}
Top view of the metallic parallelepipedic cavity ($42 \times 38.5 \times 35\ \mathrm{cm}^3 $) made reconfigurable by covering two walls with metasurfaces ($76$ physical pixels per metasurface).The EM field is probed at the locations $\vec{r}_i$ of monopole antennas $i\in [2,9]$ by measuring the transmissions between each of the latter and the monopole antenna $1$ with a vector network analyzer (VNA).}
\end{figure}

Our experimental set-up consists in an aluminum parallelepipedic cavity (see the photograph of the RC in Fig.~\ref{fig_manip} and the caption for further details). Two walls are covered with ERMs which are connected to a computer via USB. The ERMs are manufactured by GREENERWAVE  \cite{GREENWEB}.
Each metasurface consists of $76$ phase-binary pixels which are based on hybridizing two resonances \cite{kaina2014hybridized}. Each pixel can be configured electronically to control independently both tangential components of the reflected electric field by imposing a $0$ or $\pi$ phase thus leading to a total number of $152$ effective pixels per ERM. Since the design of the ERMs is based on resonant effects, the frequency band over which they stir efficiently is limited to $1\ \mathrm{GHz}$ around $5.2\ \mathrm{GHz}$. In the RC, we placed a total number of 9 monopole antennas supported by polystyrene cones of different heights. Antenna 1 is used as the emitter and the EM field is probed at the locations $\vec{r}_i$ of the remaining 8 antennas ($i\in [2,9]$) by measuring the complex transmission $S_{1i}$ between each of the latter and antenna $1$ with a vector network analyzer (VNA).

To guarantee that the RC is chaotic and therefore respecting RMT statistics  \cite{Gros2016,Gros2019,Gradoni2014,Dietz2015,Kuhl2013,Kuhl2017 }and the related spatial field uniformity \cite{Gros2014Wamot,Gros2015aem,Gros2015}, we start from a random configuration of each effective pixel. To implement a parametric stirring process with the ERMs, we successively switch a randomly chosen fixed number of pixels $\Npix$ at each step $n$. From step $n$ to $n+1$ we ensure that all pixels that have been switched at step $n$ are not modified at step $n+1$. We have performed $\Ntot=1000$ steps for $\Npix=1,4,8,16,32,48,$ and 64 starting from a different random configuration for each $\Npix$. For each step we measured 1601 frequencies from $5$\,GHz to $5.4$\,GHz with a constant frequency step of 250\,kHz. The RC including the ERMs has a quality factor $Q=356$ leading to a modal overlap $d=20.3$ \cite{Gros2016,Gros2019,Kuhl2017}, corresponding to a strong modal overlap regime close to the so-called Schroeder-Ericsson-Hill regime \cite{Schroeder1962,Ericson1963,hill2009electromagnetic}.
\begin{figure}
  \centerline{\includegraphics[width=\columnwidth]{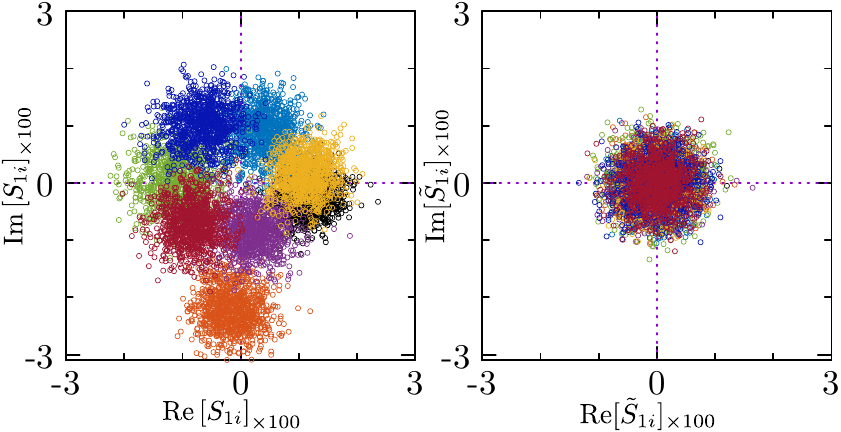}}
  \caption{\label{fig_nuage}
  Left panel: Distributions of transmissions $S_{1i}$ between antenna $1$ and antennas $i \in [2,9]$ in the complex plane at $5.2$\,GHz for 1000 random configurations of ERMs. A different color is used for each antenna. Right panel: Distributions of the corresponding centered transmissions $\tilde {S}_{1i}$.}	 
\end{figure}

In the standard, it is assumed that the complex transmission is vanishing on average $\av{S_{1i}}=0$ for each antenna $i$. This is generally not the case, e.g.\ as in our set-up where the distances between the antennas are only few wavelengths apart and the antennas are not directional. As can be seen in the left panel of Fig.~\ref{fig_nuage}, each complex antenna transmission $S_{1i}$ shows a cloud of points centered at different values $\av{S_{1i}}\ne 0$ which can be related to the short paths\cite{Gradoni2014,Yeh2010,Corona2000,Lunden2007}. These short paths not only include line-of-sight contributions but also from trajectories that are reflected by boundaries deprived of modified pixels. The points reflect the transmission at a single frequency $f=5.2$\,GHz for 1000 random configurations. Different colors correspond to different receiving antennas. %The average values vary between ... and .... 
The orange cloud is related to the receiving antenna which is the closest to the emitting one.
In section \ref{sec:nosp} we will detail the effect of these short paths on the investigated quantities. 
Meanwhile, we remove this effect in the complex transmission by subtracting its average, leading to the definition of the complex centered transmission \cite{Gradoni2014}
\begin{equation}
  \tilde {S}_{1i}= S_{1i}-\av{S_{1i}}, \label{eq:stilde}
\end{equation}
where $\av{\cdots}$ stands for configurational average.
In the right panel of Fig.~\ref{fig_nuage}, the complex centered transmissions are shown and similar fluctuations can be seen. For the characterization of antennas in RC, these centered transmissions are commonly used \cite{hol12,kro18}.
In the following section, we use this quantity to comply with the requirements of the standard.

\section{Ideal case: without short path}
\label{sec:nosp}
\begin{figure}
	\centerline{\includegraphics[width=\columnwidth]{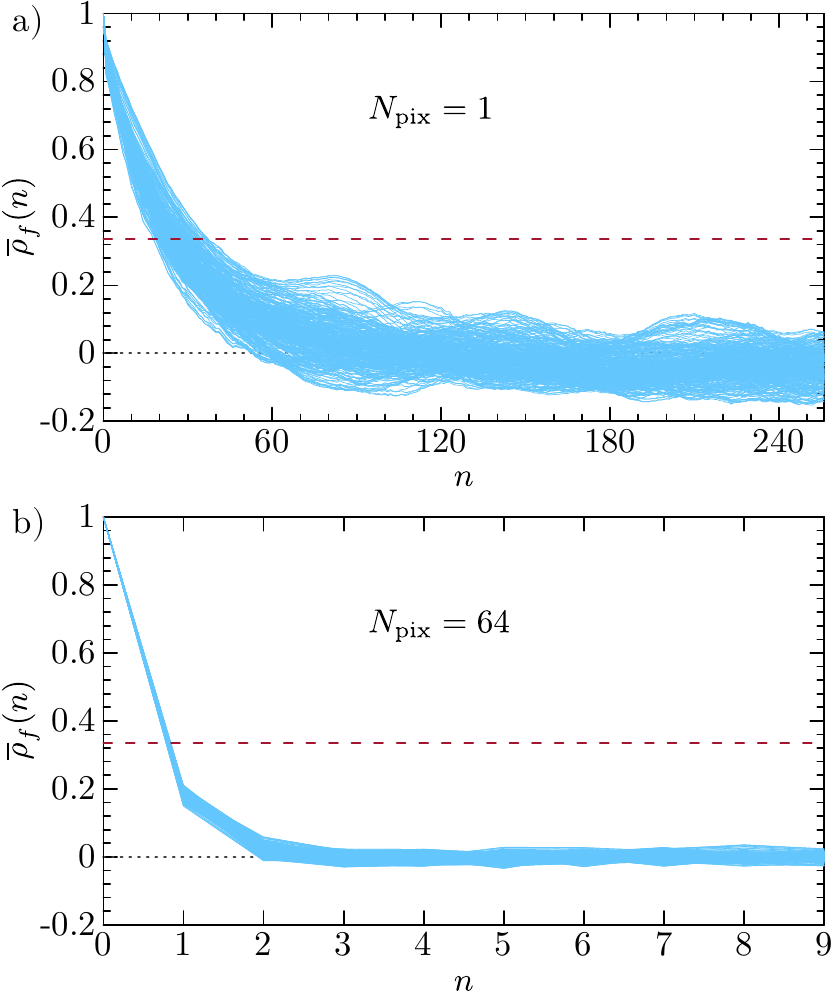}}
	\caption{ Mean autocorrelation function $\mean{\rho}_f (n)$. a) The sequence of configurations is obtained by choosing randomly $\Npix=1$ effective pixel at each step and switching its state.	The shown curves consist of discrete points linked by straight lines and correspond to different frequencies in the range {$5.0$\,--\,$5.4$\,GHz}. The horizontal dashed line corresponds to the threshold value $\rholim$ \eqref{rhol} under which the sequence is admitted to be uncorrelated according to the standard \cite{IEC_standard}. b) The sequence of configurations is obtained by choosing randomly $\Npix=64$ effective pixels at each step and switching its state.}
	\label{fig:ac}
\end{figure}

To determine the number of uncorrelated configurations, we first introduce the parametric autocorrelation function. It estimates the self-similarity of the set of $\Ntot$ data points as a function of the \emph{stir lag} $n$ and is defined for each frequency $f$ and antenna $i$ by \cite{IEC_standard} :
\begin{equation}
    \rho_{i,f}(n)=\frac{1}{\Ntot \av{x_{i,f}^2}}\sum_{l=1}^{N_\textrm{tot}} x_{i,f}(l+n) x_{i,f}(l) \label{autocorr}
\end{equation}
where
\begin{equation}
    x_{i,f}(l)=\abs{\tilde{S}_{1i,f}(l)}^2-\aver{\abs{\tilde{S}_{1i,f}(l)}^2}. \label{x}
\end{equation}
The argument $l+n$ is modulo $\Ntot$ thus periodizing the sequence $\{x_{i,f}\}$.  To reduce the fluctuations, we consider the mean autocorrelation function $\rhobar$ obtained by averaging $\rho_{i,f}$ over the 8 antenna positions $i\in[2,9]$. A conventional way to extract the number of uncorrelated configurations is given by
\begin{equation}\label{eq:Nc}
N_\textrm{c}=\frac{\Ntot}{\nlim}\,,
\end{equation}
where $\nlim$ is defined as the smallest stir lag such that the autocorrelation function $\rhobar$ goes under the threshold value proposed  by the current standard \cite{IEC_standard,Krauthauser2007}
\begin{equation}
\rholim= e^{-1} \left( 1-\frac{7.22}{\Ntot^{0.64}}\right). \label{rhol}
\end{equation}
Since $n$ is an integer, it implies that
\begin{eqnarray}
\rhobar(\nlim-1) & > \rholim \\
\rhobar(\nlim) & \leq \rholim\,.
\end{eqnarray}
Note that in the previous standard \cite{61000-4-212003}, $\rholim$ was independent of the sample size and equal to $e^{-1}$.
\begin{figure}
	\centerline{\includegraphics[width=\columnwidth]{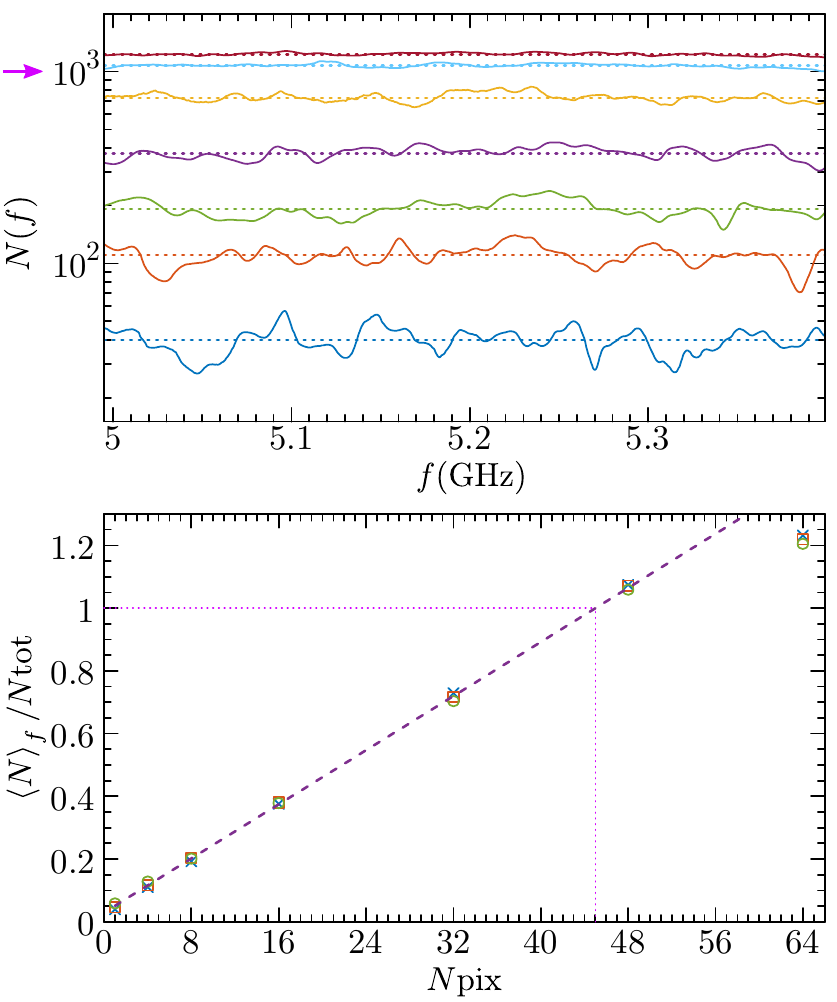}}
	\caption{Top: The continuous curves show the number of uncorrelated configurations vs frequency. In ascending order $\Npix=1,4,8,16,32,48,$ and 64. The horizontal dashed curves correspond to the frequency average $\av{N}_f$. The horizontal arrow marks the value $\Ntot=1000$. 
		Bottom: Crosses, squares and circles correspond to the values of $\av{N}_f/\Ntot$ as a function of $\Npix$ for $\Ntot=1000$, $750$, $500$, respectively. The dashed line stands for a linear fit of values $\av{N}_f/\Ntot \leq 1$ with $\Ntot=750$.
		$\av{N}_f=\Ntot$ roughly corresponds to $\Npix=45$ (dotted lines).
		\label{fig:Nf:nosp}}
\end{figure}

Fig.~\ref{fig:ac} shows $\mean{\rho}_f $ vs the stir lag $n$ for all frequencies and for the two extreme values $\Npix=1$ (a) and $64$ (b).
The horizontal dashed line corresponds to the threshold value $\rholim$. For $\Npix=1$, the mean autocorrelation function, for different frequencies, falls below $\rholim$ over a range from $n\simeq20$ to $n \simeq 40$. In case of $\Npix=64$, we find $\nlim= 1$ for all frequencies due to the discretization in the definition of $\nlim$. In both cases, for large $n$, $\rhobar$ fluctuates around zero but with a larger amplitude of fluctuations for $\Npix=1$. This can be understood by a lower number of uncorrelated samples within the total ensemble. Note that once $\Npix$ is large enough, the number of uncorrelated samples $N_\textrm{c}$ is limited by $\Ntot$. To overcome this restriction, we use a refined method which linearly interpolates the abscissa $\alim$ of the intersection of the autocorrelation function with $\rholim$ between $\nlim-1$ and $\nlim$ \cite{Oubaha2018}. We will henceforth define the number of uncorrelated configurations by
\begin{equation}
N(f)=\frac{\Ntot}{\alim} \label{eq:Nf}\,.
\end{equation}
	In the top panel of Fig.~\ref{fig:Nf:nosp}, $N(f)$ is shown for different values of $\Npix$. Each curve fluctuates around its average value $\av{N}_f$ which is constant over the presented frequency range and is indicated by a horizontal dashed line. The average value is increasing monotonously with $\Npix$, exceeding $\Ntot=1000$ (purple arrow) for $\Npix=48$ and $64$. In the bottom panel, the ratio $\av{N}_f / \Ntot$ is presented as a function of $\Npix$ for three different values of $\Ntot $. The same linear increase is found for values of $\av{N}_f / \Ntot \leq 1$ irrespectively of $\Ntot$. By using a linear regression, we can deduce the minimal amount of effective pixels to be switched at each step in order to obtain a fully uncorrelated sample. This linear regression is valid up to $\av{N}_f / \Ntot =1$ due to the refined definition~\eqref{eq:Nf}.
We would like to stress that the number of uncorrelated configurations seems to be virtually unlimited whereas, in the case of mechanical stirring
at such a low frequency, it would be well below $100$.

\begin{figure}
  \centerline{\includegraphics[width=\columnwidth]{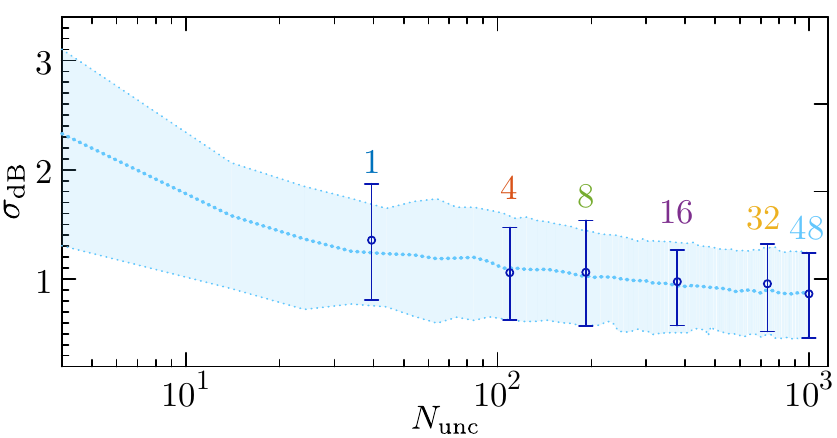}}
  \caption{\label{fig:sdb:nosp}
  Confidence interval (95\,\%) of $\sigma_\textrm{dB}$ as a function of the effective number of uncorrelated configurations $\Nunc$. Circles with bars give the mean value $\av{\sigma_\textrm{dB}}_f$ and the associated confidence interval for the values of $\Npix$ indicated above the bars. For each value of $\Npix$ (except $48$), $\Nunc$ is smaller than $\Ntot$ (see Fig.~\ref{fig:Nf:nosp}). The thick dotted line and the shaded blue area give the mean value $\av{\sigma_\textrm{dB}}_f$ and the associated confidence interval when $\Ntot=\Nunc$.}
\end{figure}

 We now consider the uniformity criterion of the standard \cite{IEC_standard}, based on the so-called $\sdb$ describing the fluctuations of the maxima of the field amplitude measured at least at eight locations:
 \begin{equation}\label{eq:sdb}
 \sdb(f)=20 \log_{10}\left(1+\frac{\sigma_{\textrm{max}}}{\langle\modul{E_i}{}_{\textrm{max}}\rangle_8}\right)
 \end{equation}
At each of these locations, for the $\Ntot$ configurations of the ERMs, one extracts the amplitude of the field component $\modul{E_i}$ and one keeps the maximum value $\modul{E_i}{}_{\textrm{max}}$. One then computes the average and the standard deviation $\sigma_{\textrm{max}}$ over the 8 values of $\modul{E_i}{}_\textrm{max}$.
In our case, the field amplitude is deduced from $\tilde{S}_{1i}$ via \cite{Gros2015}
\begin{equation}
E_i= \frac{\tilde{S}_{1i}(f)}{\kappa_1\kappa_i}\label{eq:Ei},
\end{equation}
where $\kappa_i$ is the coupling constant of the antenna $i$ at the investigated frequency \cite{Kuhl2013,Kober2010,Kuhl2017,Fyodorov2005}. Since each $\kappa_i$ is constant in our case, one can evaluate \eqref{eq:sdb} directly using $\tilde{S}_{1i}$ instead of $E_i$.

More specifically, we are interested in the statistical behavior of $\sdb(f)$ with respect to the effective number of uncorrelated configurations
 \begin{equation}
\Nunc = min(\av{N}_f,\Ntot) \label{eq:nunc}
 \end{equation}
among the $\Ntot$ configurations of the ERMs. Here, as in \cite{Lemoine2009,CEM2016}, we address the tolerance requirements of the standard regarding the 3\,dB threshold level for $\sdb$. In Fig.~\ref{fig:sdb:nosp}, we show the mean values $\av{\sigma_\textrm{dB}}_f$ (blue circles) and their 95\,\% confidence interval (vertical bars) for all $\Npix$ except for $\Npix=64$ which has the same $\Nunc$ as $\Npix=48$. The thick blue dotted line is the average value $\av{\sigma_\textrm{dB}}_f$ taking into account only samples of size $\Ntot$ reduced to $\Nunc$. These samples are drawn from the measurements associated to $\Npix=48$ which ensure that the samples are fully uncorrelated. The shaded blue area is the corresponding 95\,\% confidence interval. Comparing the vertical bars with the shaded blue area it is evident that the statistical behavior of $\sdb$ is only characterized by the number of uncorrelated configurations. We would like to emphasize that the confidence interval always lies below the threshold of 3\,dB apart from a slight
excess for $\Nunc=4$.

This statistical behavior can be predicted via RMT due to the chaotic character of the RC induced by the presence of the ERMs \cite{Gros2019}.
\section{RMT predictions }
 \begin{figure}
 	\centerline{\includegraphics[width=\columnwidth]{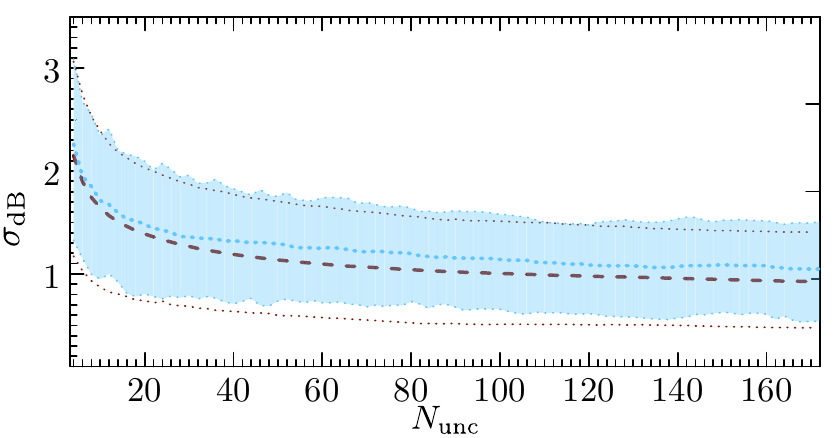}}
 	\caption{Comparison of the experimental results shown in Fig.~\ref{fig:sdb:nosp} (blue curve and shaded area) with the Monte-Carlo simulation (brown dashed and dotted curves) using the theoretical prediction \eqref{eq:Pnini:int:rep}}
 	\label{fig:rmt}
 \end{figure}
In the framework of RMT description of the universal statistical behavior of chaotic cavities, the real and imaginary parts of each component of the field are not identically distributed \cite{Gros2016}, contrary to Hill's assumptions \cite{hill2009electromagnetic}. For a given configuration of an ideally chaotic cavity, they still are independently Gaussian distributed, but with different variances\cite{Gros2019}. The ensuing distribution of the modulus of any component of the field $\modul{E_i}$ is therefore not a Rayleigh distribution \cite{Lemoine2009} but depends on a single parameter $\rho$, called the \emph{phase rigidity}\footnote{Note that for $\modul{\rho} \rightarrow 1$ the system tends to be closed, a situation corresponding to non-overlapping resonances, whereas $\modul{\rho} \rightarrow 0$ corresponds to the asymptotic limit of Hill's assumptions.} \cite{Gros2016}.
The main steps leading to the RMT distribution of the normalized field amplitude of the Cartesian component $\Ec_i=\abs{E_i}/\av{\abs{E_i}^2}^{1/2}$ deduced from a statistical ensemble resulting from stirring are given in \cite{Gros2016,Gros2015,CRC_review,Gros2019}.
We here only recall the final RMT prediction which reads
\begin{align}\label{eq:Pnini:int:rep}
  P_a(\Ec_i)=\int_0^1 P_\rho (\rho) P(\Ec_i;\rho) d\rho\,,
\end{align}
where
\begin{equation}\label{eq:Pnini}
  P(\Ec_i;\rho)=\frac{2 \Ec_i}{\sqrt{1-\modulc{\rho}}} \exp\left[ -\frac{\Ec_i^2}{1-\modulc{\rho}} \right] \textrm{I}_0\left[\frac{\modul{\rho}\Ec_i^2}{1-\modulc{\rho}} \right]
\end{equation}
is derived from the Pnini and Shapiro distribution \cite{Pnini1996,Gros2016,Kim2005} and $ P_\rho$ is the phase rigidity distribution.
Preliminary investigations, based on numerical simulations of the Random Matrix model described in \cite{Gros2016}, show that $P_\rho(\rho)$ depends only on the mean modal overlap $d$.
An ansatz was proposed in \cite{Gros2015} to determine $P_\rho(\rho)$ from the only knowledge of $d$. This Ansatz reads:
\begin{equation}
P_\rho^W=\frac{2B\exp[-2B\rho/(1-\rho)]}{(1-\rho)^2}
\end{equation}
where the parameter $B$ has a smooth $d$-dependence \cite{Gros2015} numerically deduced from our RMT model presented in \cite{Gros2016}.
Originally in \cite{Gros2015}, the empirical estimation of $B(d)$ was limited to $d\le 1$.
Currently, $B(d)$ has been extended to larger values of $d$ and is given by \cite{Gros2019,CRC_review}
\begin{equation}
B(d) = \frac{ad^2}{1 + bd + cd^2}\,,
\end{equation}
with $a=0.50\pm 0.02$, $b = 1.35 \pm 0.03$ and $c = 0.30 \pm 0.02$ \cite{CRC_review}.

Using the above distribution for the normalized field amplitude, we performed a Monte-Carlo simulation based on \eqref{eq:Pnini:int:rep} with 4096 estimations of $\sdb$. Each simulation provides one estimation of $\sdb$, from 8 estimations of the maximum among $\Nunc$ values. The value of the experimental modal overlap $d=20$ was used for the Monte-Carlo simulation.
The predicted results for the 95\,\% confidence interval and the mean value of $\sdb$ are shown in Fig.~\ref{fig:rmt}. A comparison with the experimental results of Fig.~\ref{fig:sdb:nosp} is presented. A fair agreement is obtained save for a slight excess of the experimental results more likely related to the presence of spatial correlations induced by the proximity between receiving antennas in our set-up. A larger deviation would be observed if we relied on Hill's assumptions where the underlying distribution of the field amplitudes is the Rayleigh function \cite{Lemoine2009}.

\section{Effects of short paths}
\label{sec:short}

\begin{figure}
	\centerline{\includegraphics[width=\columnwidth]{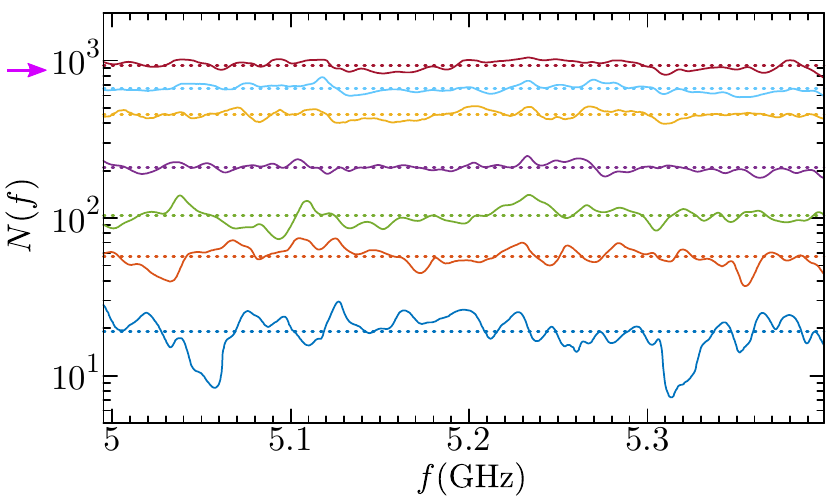}}
	\caption{Same as the top panel of Fig.~\ref{fig:Nf:nosp} where $S_{1i}$ is used instead of $\tilde{S}_{1i}$ in \eqref{x} thus including the contributions of short paths. In ascending order $\Npix=1,4,8,16,32,48,$ and 64.}
	\label{fig:Nf:wsp}
\end{figure}

We now investigate the effects of short paths on the above studied quantities, namely the number of uncorrelated configurations $N(f)$ and $\sdb$.
In \eqref{x}, \eqref{eq:Ei}, the centered complex transmission $\tilde{S}_{1i}$ is replaced by the complex transmission $S_{1i}$ itself.
The ensuing curves for $N(f)$ are shown in Fig.~\ref{fig:Nf:wsp} to be compared with Fig.~\ref{fig:Nf:nosp}(top).
It appears that, for each value of $\Npix$, the frequency average $\av{N}_f$ is systematically reduced with respect to the value obtained when the centered complex transmission is used.
In the present case, the average value does not exceed $\Ntot=1000$ even for $\Npix=64$ thus leading to a reduced maximum value of $\Nunc$.

\begin{figure}
	\centerline{\includegraphics{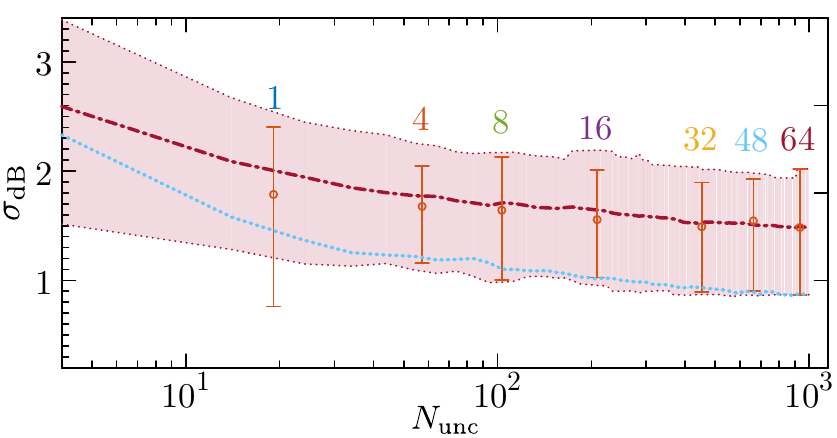}}
	\caption{Confidence interval (95\,\%) of $\sdb$ as a function of the effective number of uncorrelated configurations $\Nunc$ including the contributions of short paths. Red circles with bars give the mean value $\av{\sdb}_f$ and the associated confidence interval for the values of $\Npix$ indicated above the bars. The thick dash-dotted red line and the shaded light red area give the mean value $\av{\sdb}_f$ and the associated confidence interval when $\Ntot=\Nunc$. The dotted blue line of Fig.~\ref{fig:sdb:nosp} is shown for the sake of comparison.}
	\label{fig:sdb:wsp}
\end{figure}

In Fig.~\ref{fig:sdb:wsp} we present the mean value $\av{\sdb}_f$ and its 95\,\% confidence interval to be compared with Fig.~\ref{fig:sdb:nosp}.
In this case the shaded light area is calculated drawing the samples from the configurations with $\Npix=64$.
Again the mean value is decreasing monotonically but this time is above the mean value without short paths (see blue dashed lines).
For small $\Nunc \le 8$ the confidence interval exceeds the 3\,dB threshold which in case of no short paths was only exceeded for $\Nunc\le 4$.
The predictions from RMT are not respected anymore due to the non-universal contributions of short paths.
For testing wireless devices \cite{hol06,Remley2016,Reis2018} and characterization of antennas \cite{hol12,kro18} it is crucial to be able to remove these contributions.

\section{Conclusion}
We proposed and experimentally verified a new method to stir the EM field in reverberation chambers. This method relying on ERMs is efficient, much faster than the conventional mechanical mode stirring, and above all, it allows us to control the number of uncorrelated configurations at will. Furthermore, since the ERMs make the RC chaotic, the statistical field uniformity required by many applications is automatically ensured and RMT can be used to describe its statistical behavior even when the Hill regime is not reached. We have also shown that the number of uncorrelated configurations is the parameter that truly controls the $\sdb$ confidence interval for a given modal overlap. Therefore, increasing the sample size above the number of uncorrelated configurations has no effective impact on this confidence interval. In addition, we have investigated the effects of short paths. The latter introduce extra correlations, which automatically reduce the number of uncorrelated configurations and accordingly increase $\sdb$. 

Finally, from a practical point of view, we would like to emphasize that stirring via the inclusion of the ERMs allows to perform a huge number of uncorrelated configurations.
Therefore characterizations which rely on the statistical behavior of the field in RCs are more accurate.
This accuracy can be a strong requirement in many applications such as radar cross section \cite{Lerosey2007,Baba2012,Reis2018}, material absorption cross section \cite{Amador2011,Flintoft2016,Gradoni2012}, MIMO and antenna characterization \cite{kil04,hol12,kro18,che18,Remley2016,davy2,Rosengren2005,mimo,Gradoni2014_mimo}, and Ricean radio environment for testing wireless devices \cite{hol06,Lemoine2011,Savin2017,Sanchez-Heredia2011}.

\section*{Acknowledgment}
The authors acknowledge funding from the French ``Minist\`{e}re des Arm\'{e}es, Direction G\'{e}n\'{e}rale de l'Armement'' and  the European Commission for financial support through the H2020 program by the Open Future Emerging Technology ``NEMF21'' Project 664828.

\bibliographystyle{IEEEtran}
%\bibliography{Biblio}
% Generated by IEEEtran.bst, version: 1.14 (2015/08/26)

\end{document}